\documentclass[10pt,a4]{article}
\usepackage[in]{fullpage}
\usepackage{amsmath,amssymb}
\usepackage{cite}
\usepackage{xspace}
\usepackage{psfrag,graphicx}
\usepackage{lineno}

\long\def\symbolfootnote[#1]#2{\begingroup%
\def\thefootnote{\fnsymbol{footnote}}\footnote[#1]{#2}\endgroup} 

\newcommand{\xmax}{\ensuremath{X_\mathrm{max}}}
\newcommand{\xmaxv}{\ensuremath{X_\mathrm{max}^\mathrm{v}}}
\newcommand{\gsm}{g/cm${}^2$}
\newcommand{\dgv}{$DG^\mathrm{V}$}

\newcommand{\sigrat}{$S_\mu/S_\mathrm{EM}$}
\newcommand{\denrat}{$D_\mu/D_\mathrm{e}$}
\newcommand{\gss}{0.75}
\newcommand{\logen}{$\log10(E)$[eV]=}

\psfrag{(Sem[3][3]+Sgm[3][3])}[r][r][\gss]{$S_\mathrm{EM}$, VEM}
\psfrag{(870\255Xmax)/cos(Zth)}[r][r][\gss]{$DG$, \gsm}
\psfrag{870\255Xmax}[r][r][\gss]{\dgv, \gsm}
\psfrag{(Smu[3][3])}[r][r][\gss]{$S_\mu$, VEM}
\psfrag{Smu[3][3]/(Sem[3][3]+Sgm[3][3])}[r][r][\gss]{\sigrat}
\psfrag{FitMu}[r][r][\gss]{$(S_\mu^\mathrm{MC}-S_\mu^\mathrm{fit})/S_\mu^\mathrm{MC}$, \%}
\psfrag{Smu[0][3]/(Sem[0][3]+Sgm[0][3])}[r][r][\gss]{$S_\mu^{200}/S_\mathrm{EM}^{200}$}
\psfrag{Smu[4][3]/(Sem[4][3]+Sgm[4][3])}[r][r][\gss]{$S_\mu^{1500}/S_\mathrm{EM}^{1500}$}
\psfrag{Xmax}[r][r][\gss]{\xmaxv, \gsm}
\psfrag{DFitMu}[r][r][\gss]{$(D_\mu^\mathrm{MC}-D_\mu^\mathrm{fit})/D_\mu^\mathrm{MC}$, \%}
\psfrag{(Dmu[3][3]/Dem[3][3])}[r][r][\gss]{\denrat}
\psfrag{Showers}[r][r][\gss]{Showers}

\begin{document}
\mbox{}\\[.7cm]

\begin{center}
{\large\sc Precise determination of muon shower content\\[0.08cm] from shower universality property}\\[0.7cm]

A.~Yushkov${}^1$\symbolfootnote[1]{Corresponding author; e-mail:
  yushkov@na.infn.it}, M.~Ambrosio${}^1$, C.~Aramo${}^1$,
F.~Guarino${}^{1,2}$, D.~D'Urso${}^1$, L.~Valore${}^1$\\[0.3cm]

${}^1$INFN Sezione di Napoli, 80125, via Cintia, Napoli,
  Italia\\
${}^2$Universit\`{a} di Napoli ``Federico~II'', 80125, via Cintia, Napoli,
  Italia\\[0.5cm]
\end{center}


\begin{abstract}
It is shown, that highly accurate estimation of muon shower content
can be performed on the basis of knowledge of only vertical depth of
shower maximum \xmaxv\ and total signal in ground detector. The
estimate is almost independent on primary energy and particle type and
on zenith angle. The study is performed for 21500 showers, generated
with CORSIKA~6.204 from spectrum $E^{-1}$ in the energy range
$\log10(E)$ [eV]=18.5--20 and uniformly in $\cos^2{\theta}$ in zenith
angle interval $\theta=0^\circ-65^\circ$ for QGSJET~II/Fluka
interaction models.
\end{abstract}

Muon shower content is the key parameter for studies of primary mass
composition and test of hadronic interaction models. Using
universality property of extensive air showers (EAS)
development~\cite{giller_univers2005,nerling_univers2005,gora_univers2006,ave_icrc30_univers,lipari_univers2008,lafebre_univers2009}
we propose a simple and precise method to determine muon shower
content from vertical depth of shower maximum and total signal (signal
in water tanks or in scintillator detector). The study is performed
for 21500 showers, generated with CORSIKA~6.204~\cite{corsika} from
spectrum $E^{-1}$ in the energy range $\log10(E)$ [eV]=18.5--20 (with
different statistics in 3 energy bins 18.5--19.0, 19.0--19.5 and
19.5--20.0) and uniformly in $\cos^2{\theta}$ in zenith angle interval
$\theta=0^\circ-65^\circ$ for
QGSJET~II~\cite{qgsjetii,qgsjetiia,qgsjetiib}/Fluka~\cite{fluka1,fluka2}
interaction models. Electromagnetic (EM) component thinning was set to
$10^{-6}$, the observation level was at 870~\gsm. All longitudinal
showers characteristics and charged particles density were taken
directly from CORSIKA output files. The expected signal in Auger-like
tanks was calculated according to the procedure described
in~\cite{billoir_sampl_2008,ave_munum_2007} with the use of the same
GEANT~4 lookup tables as in~\cite{ave_munum_2007}.

\begin{figure}
\includegraphics[width=0.49\textwidth]{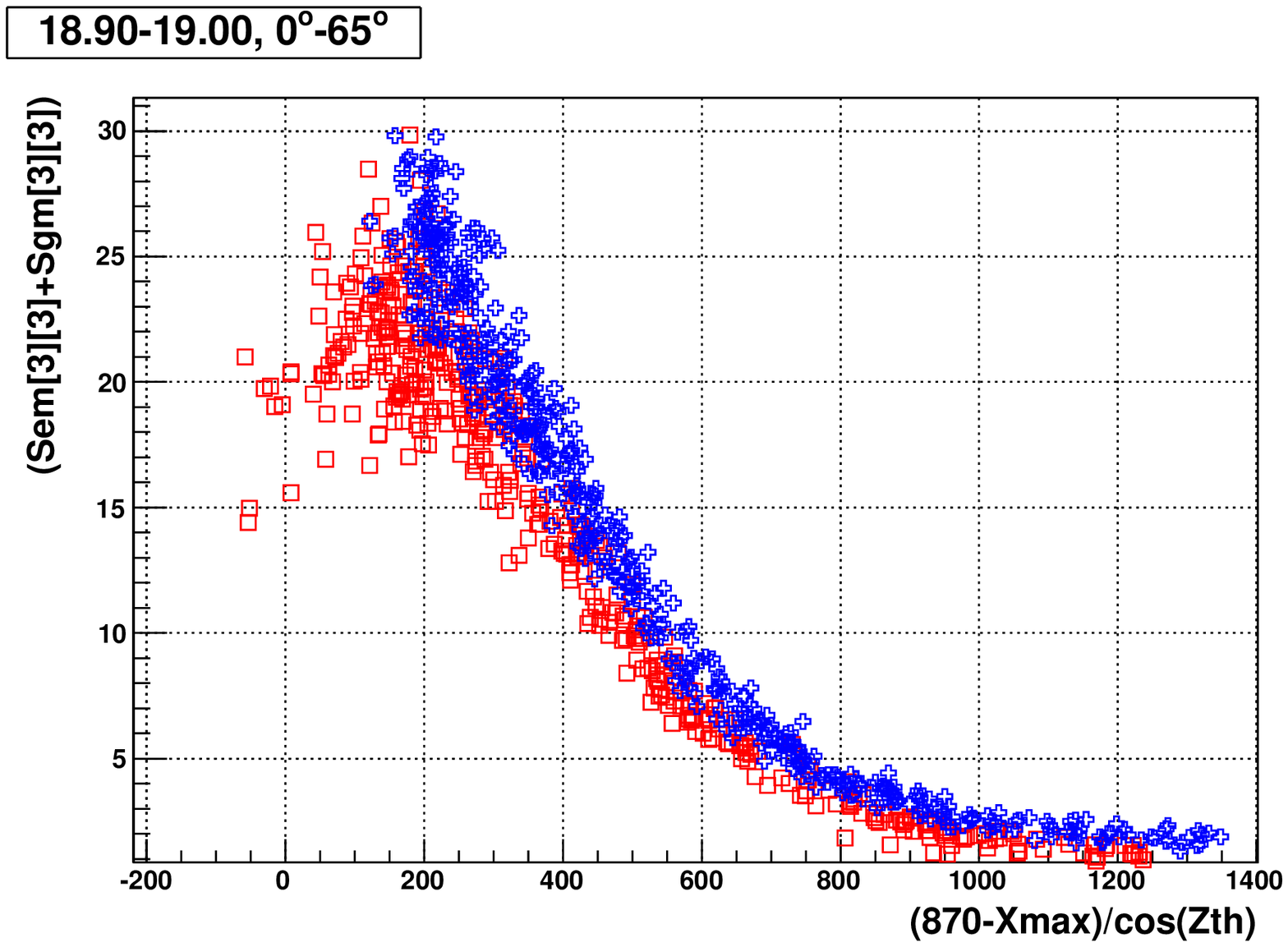}
\includegraphics[width=0.49\textwidth]{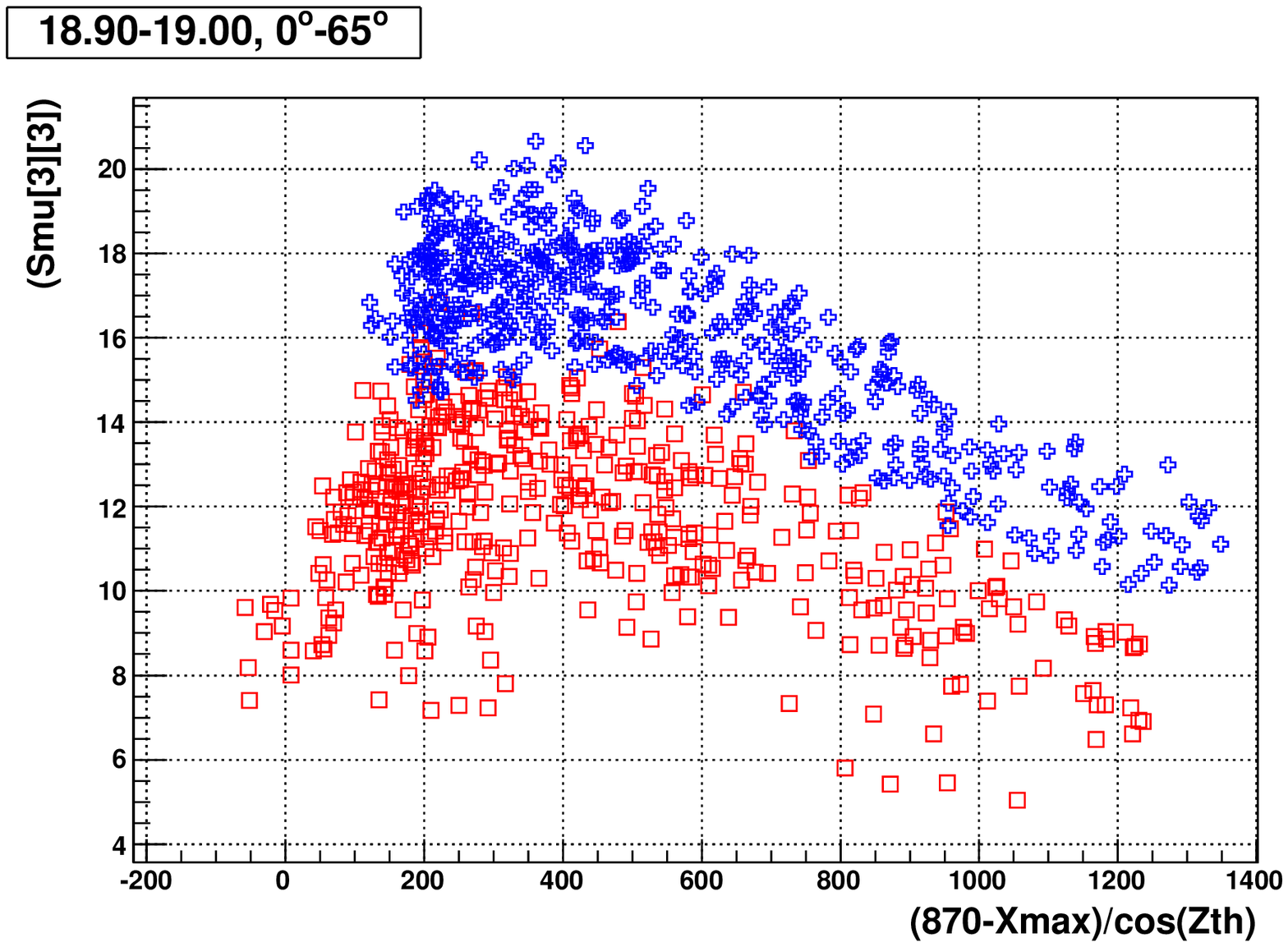}
\caption{EM and muon signals from proton (red squares) and iron (blue
  crosses) in water Cherenkov tanks at 1000~m vs slant distance from
  shower maximum to the ground $DG$ in \logen18.9--19.0 energy bin}
\label{DGSlant}
\end{figure}

\begin{figure}
\includegraphics[width=0.49\textwidth]{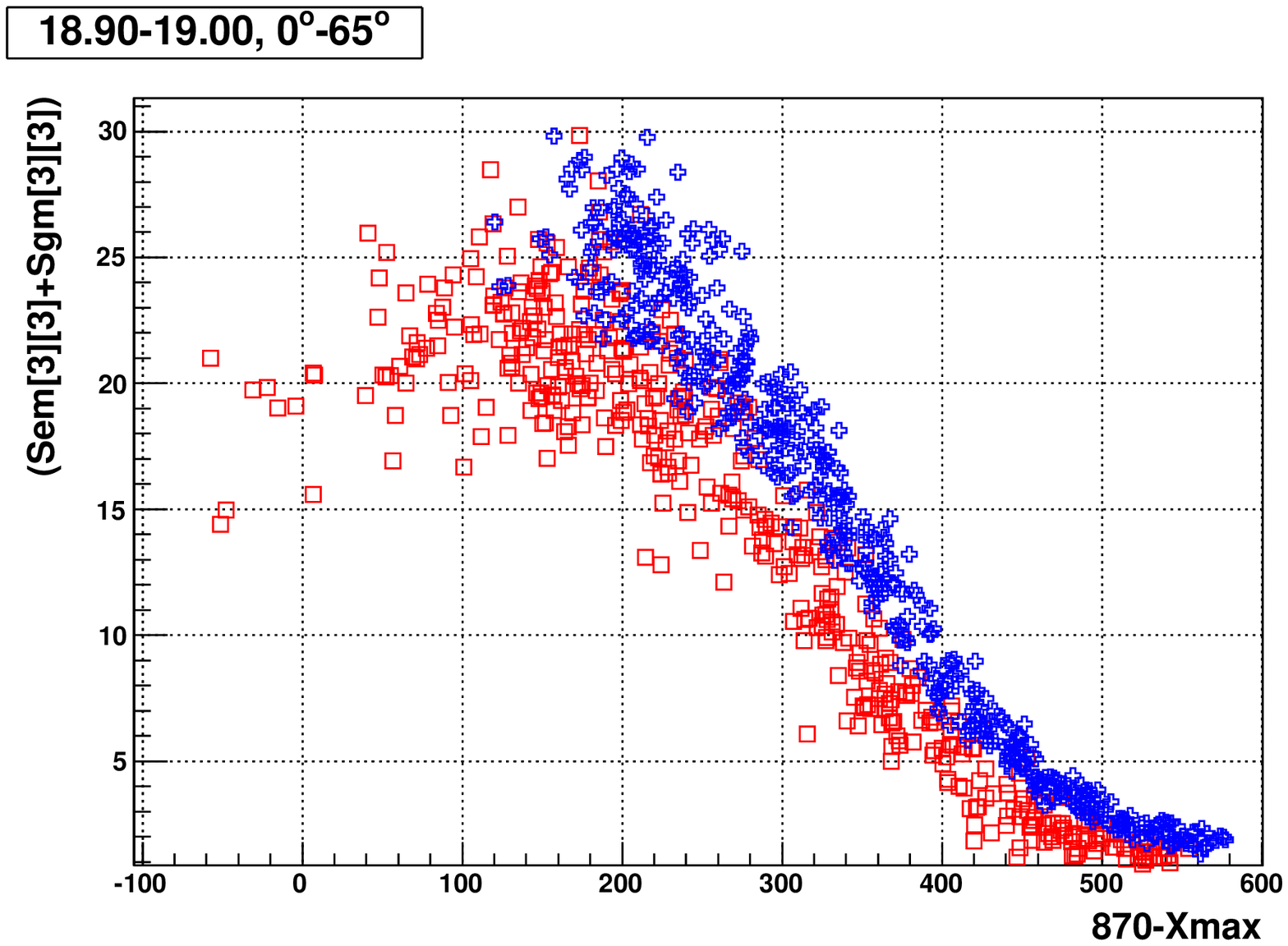}
\includegraphics[width=0.49\textwidth]{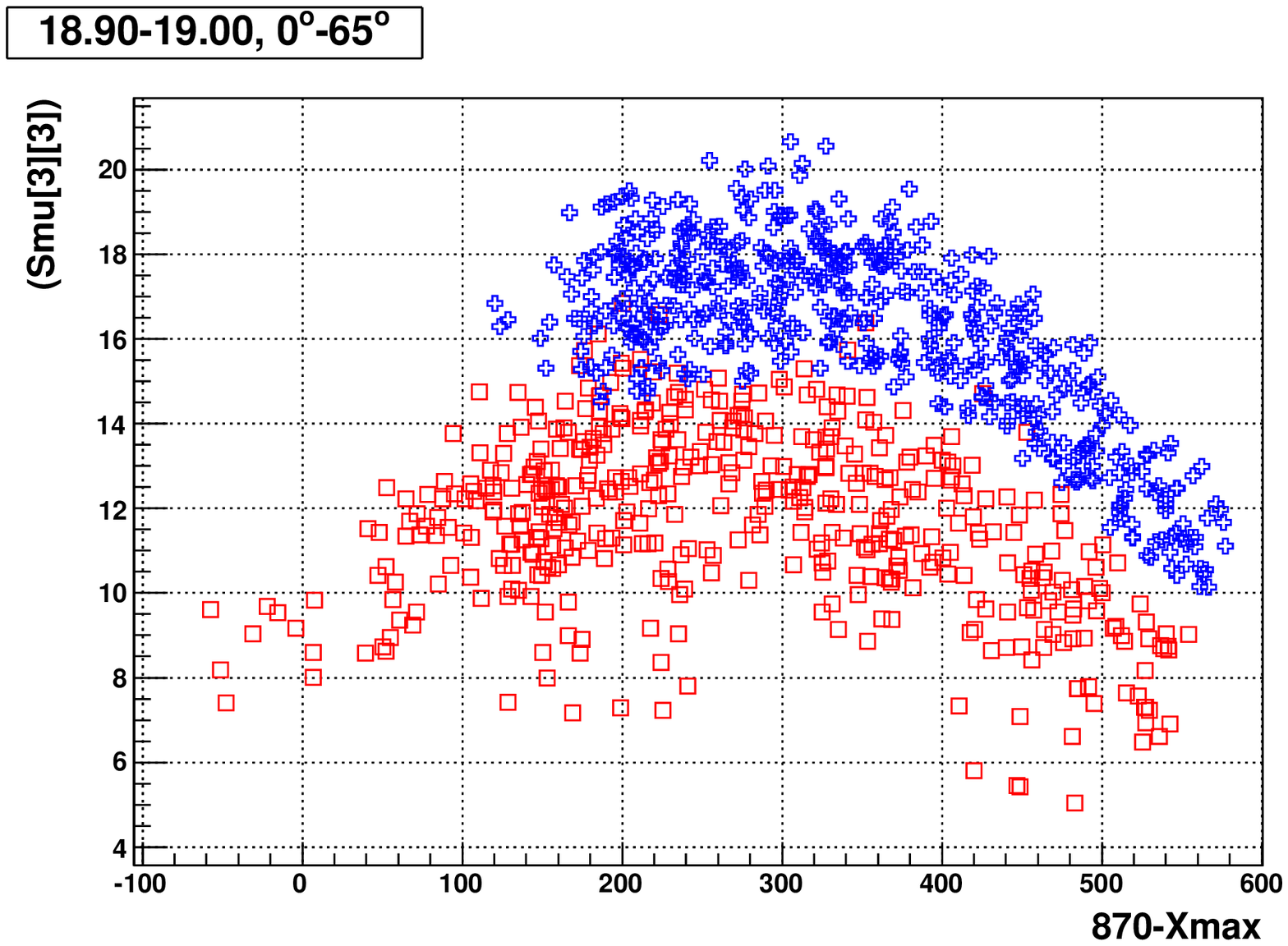}
\caption{EM and muon signals from proton (red squares) and iron (blue
  crosses) in water Cherenkov tanks at 1000~m vs vertical distance
  from shower maximum to the ground \dgv in \logen18.9--19.0 energy bin}
\label{DGVert}
\end{figure}

\begin{figure}
\includegraphics[width=0.49\textwidth]{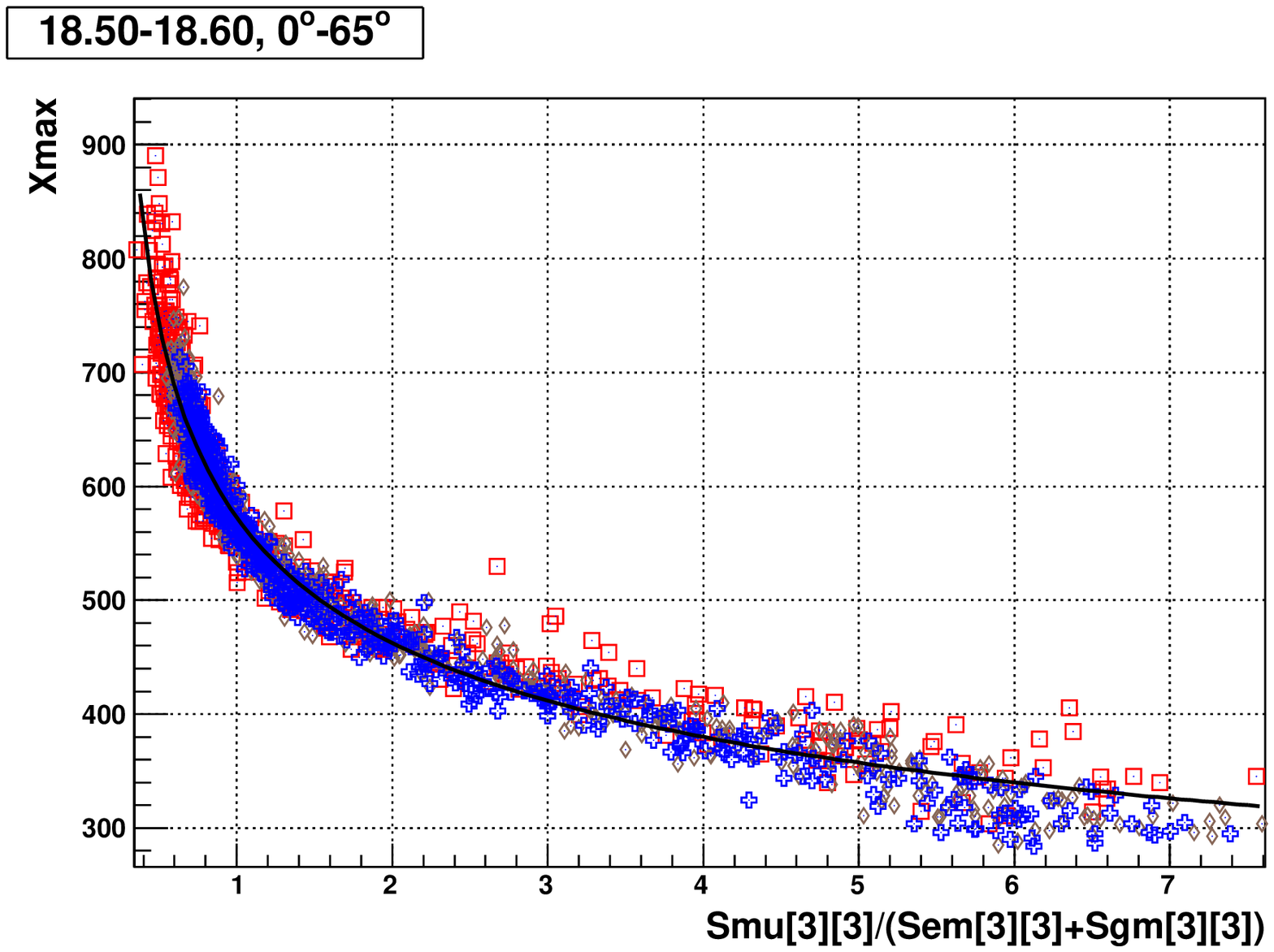}
\includegraphics[width=0.49\textwidth]{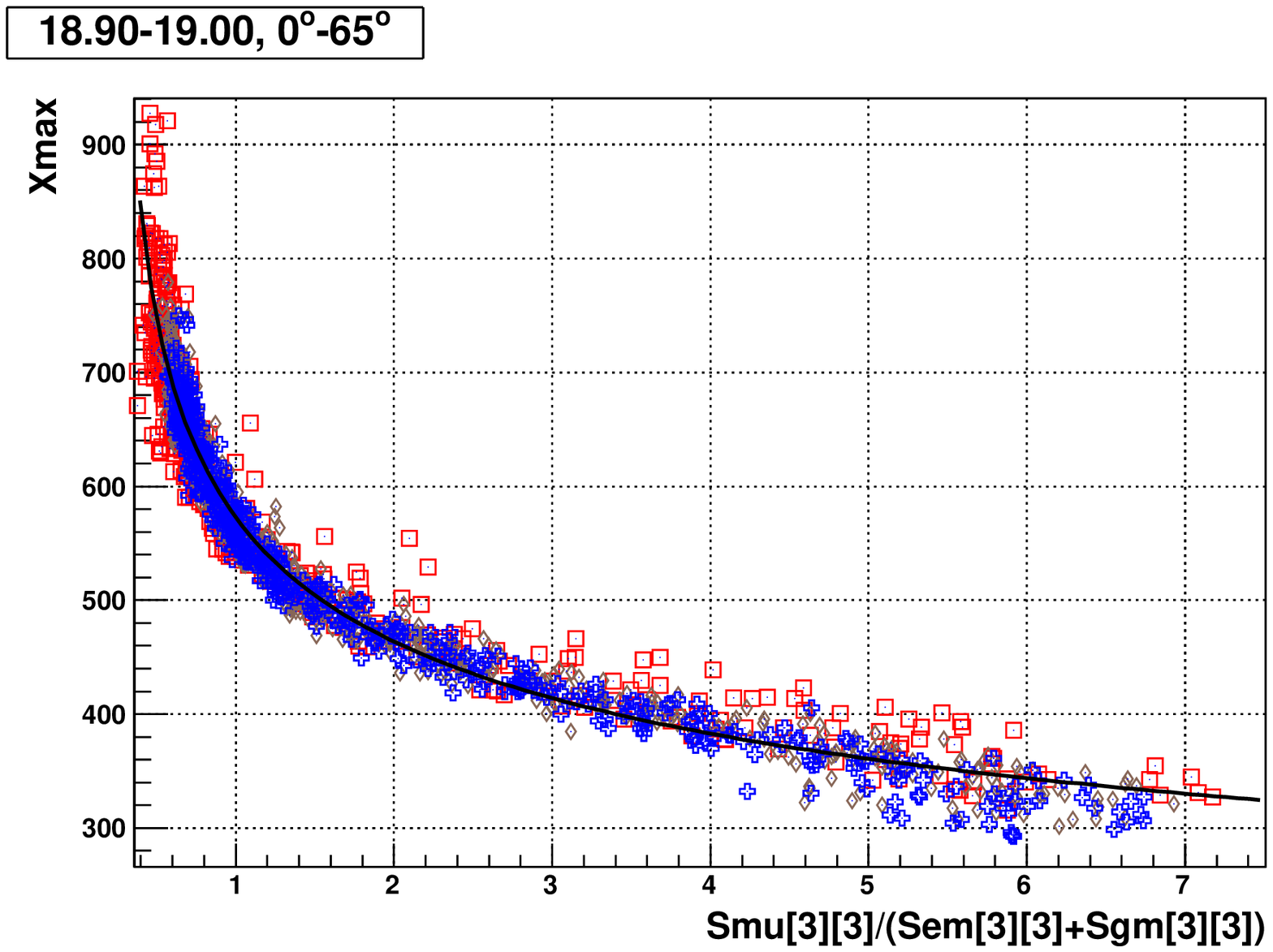}
\includegraphics[width=0.49\textwidth]{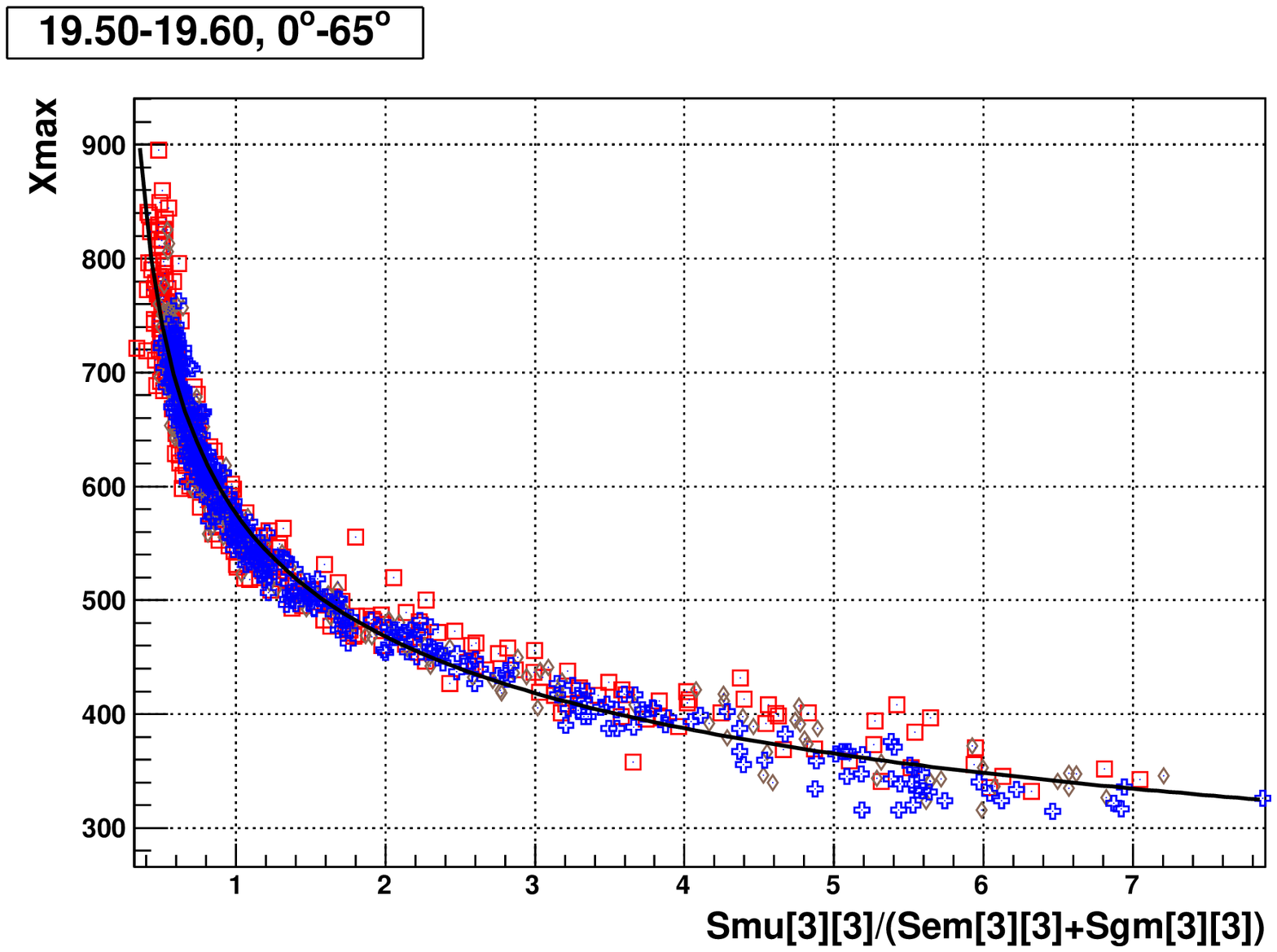}
\includegraphics[width=0.49\textwidth]{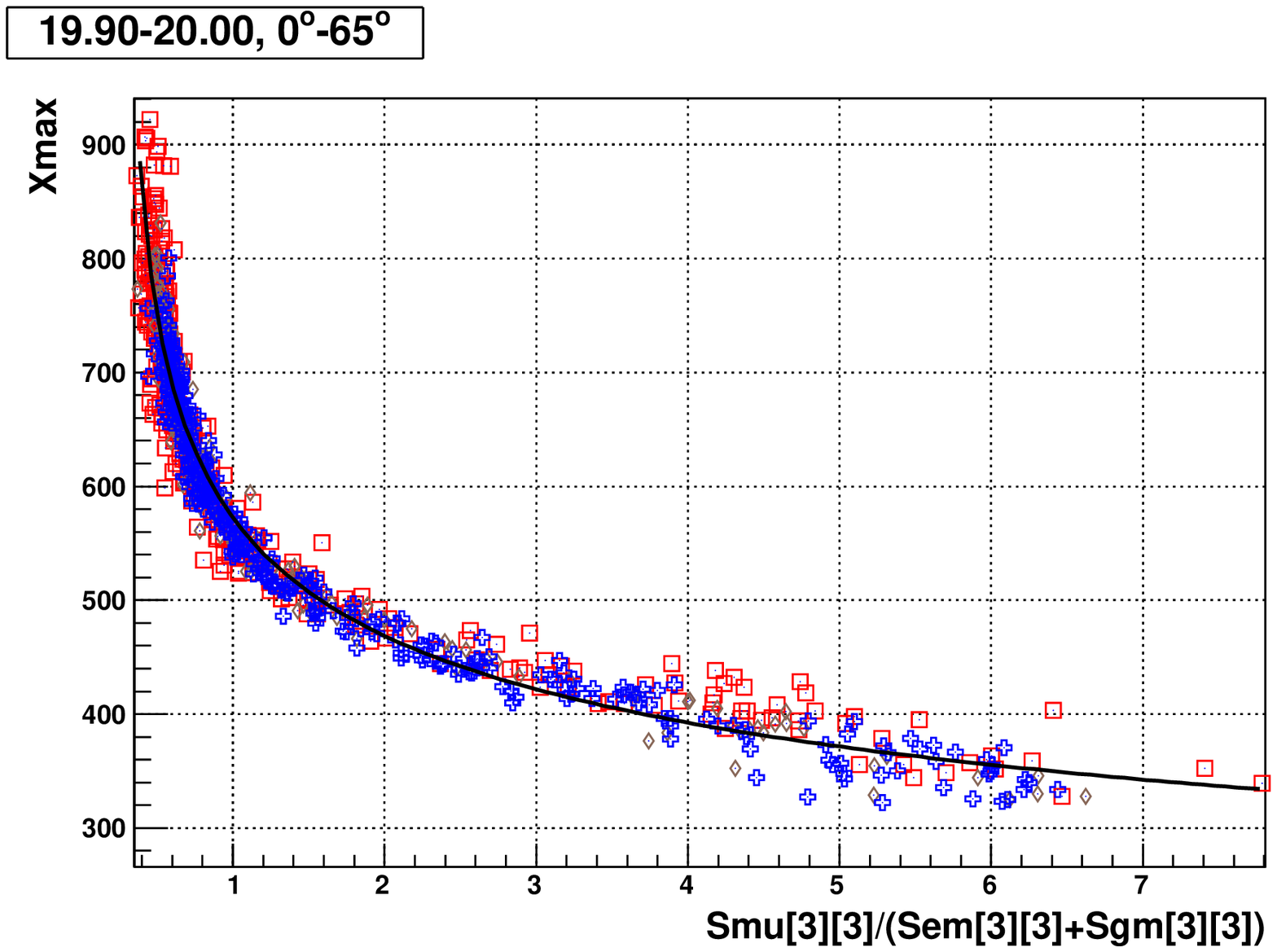}
\caption{Ratio of signals in water Cherenkov tanks \sigrat\ at 1000~m
  vs vertical depth of shower maximum \xmaxv\ in four energy bins. Protons~---~red squares,
  oxygen~---~brown diamonds, iron~---~blue  crosses}
\label{muemxmax}
\end{figure}

\begin{figure}
\includegraphics[width=0.49\textwidth]{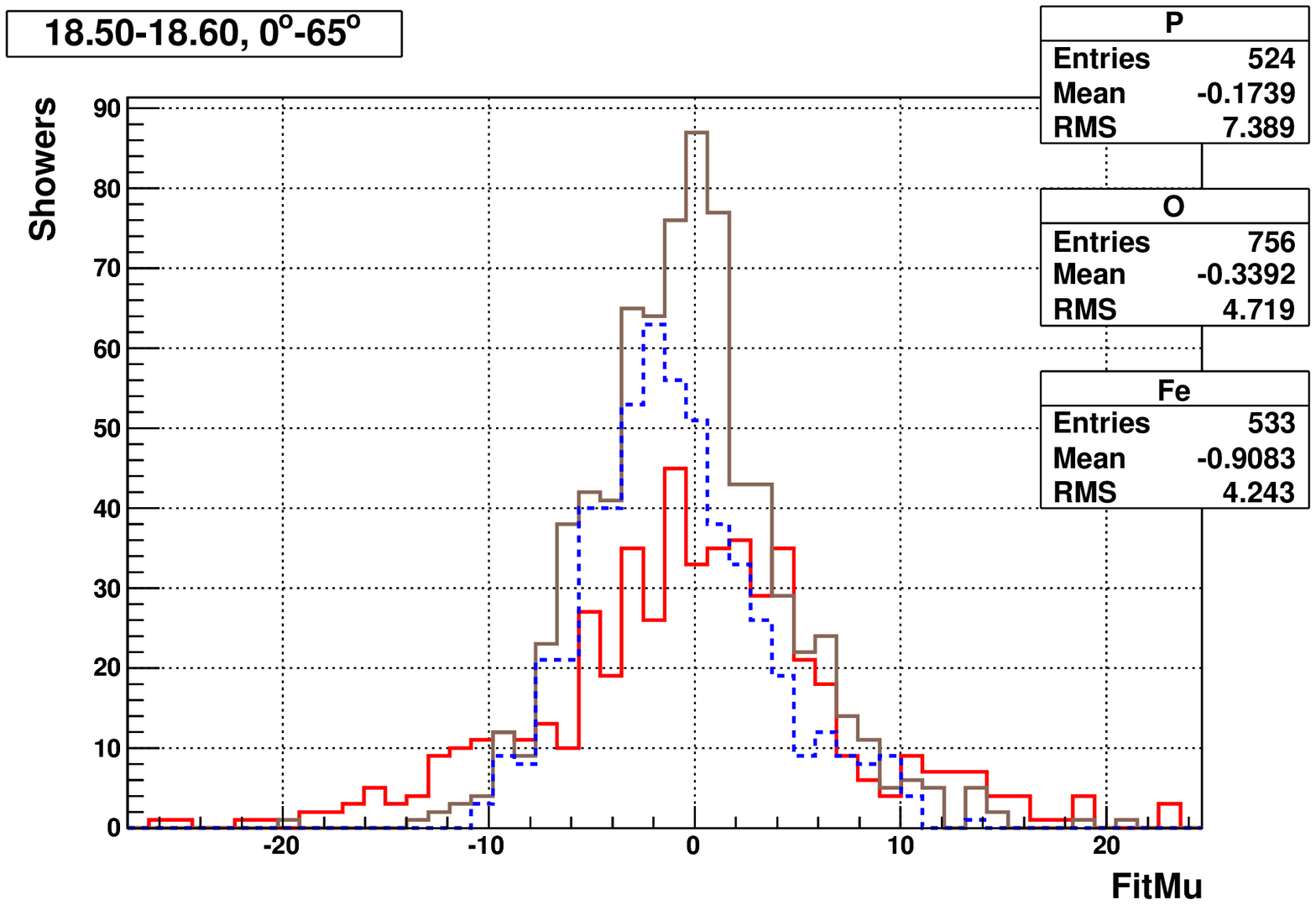}
\includegraphics[width=0.49\textwidth]{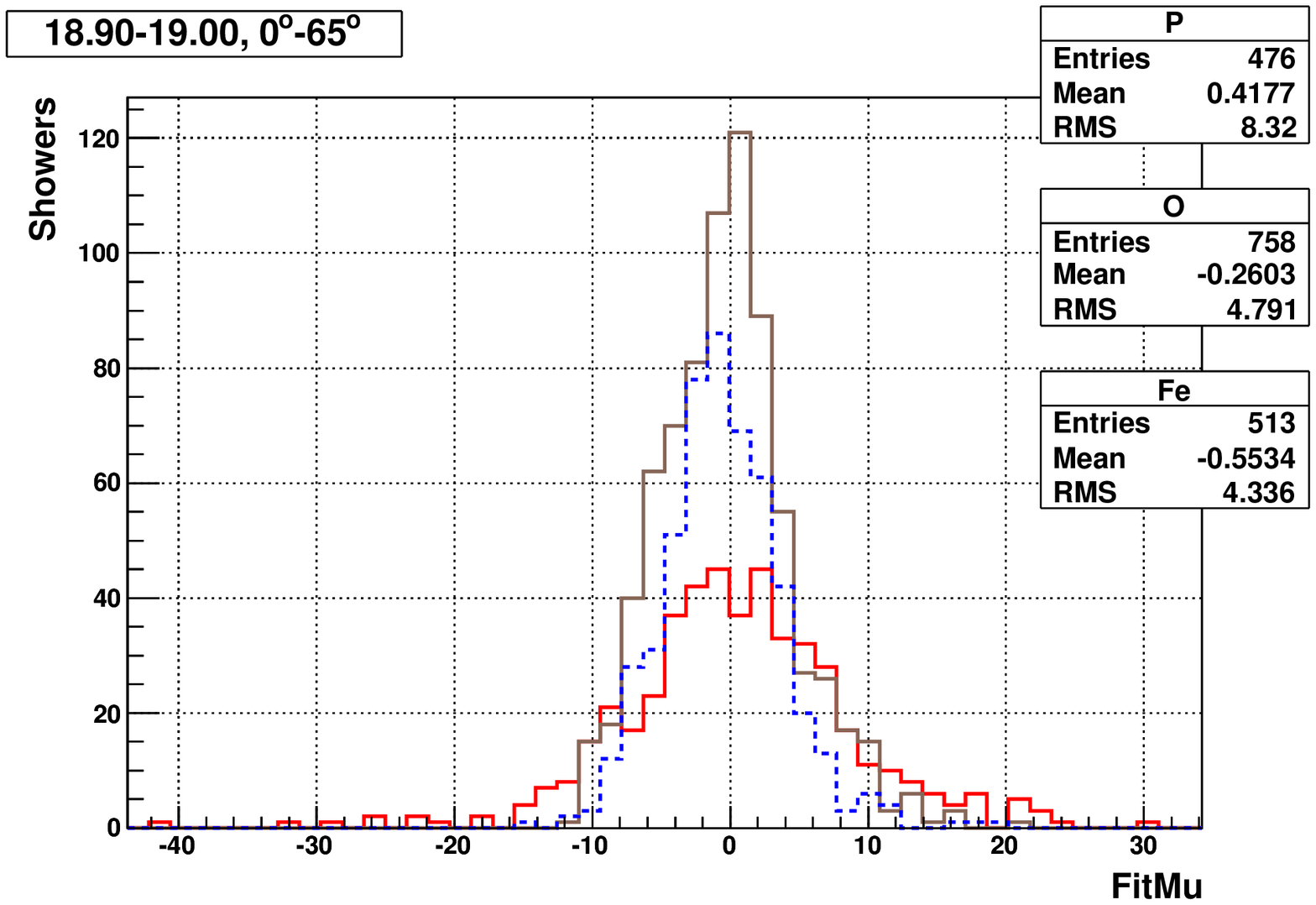}
\includegraphics[width=0.49\textwidth]{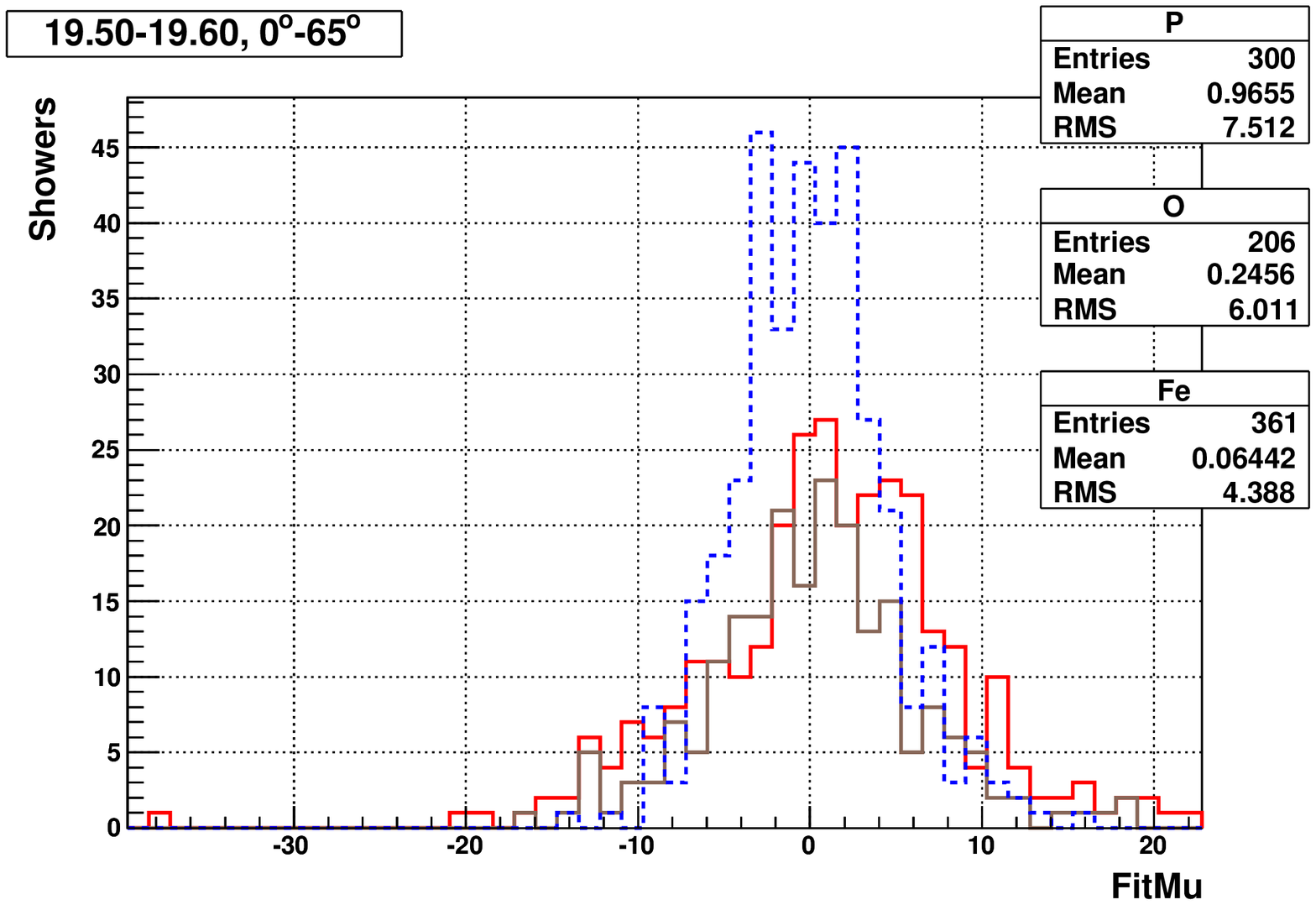}
\includegraphics[width=0.49\textwidth]{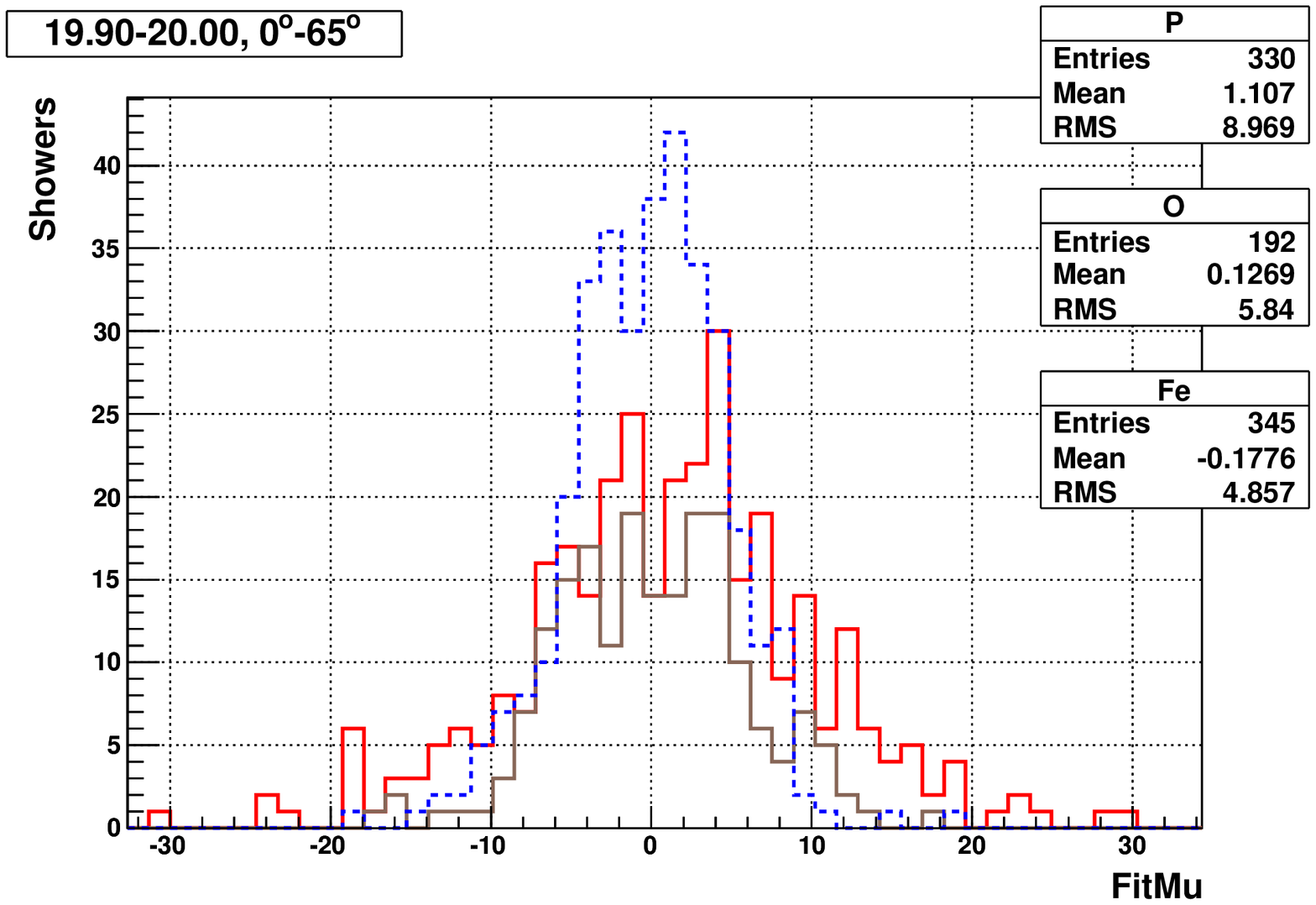}
\caption{Distributions of relative difference between MC simulated muon
  signals in Cherenkov water tanks $S_\mu^\mathrm{MC}$ and muon signals derived from the fit
  $S_\mu^\mathrm{fit}$ at 1000~m. Protons~---~red line,
  oxygen~---~brown line, iron~---~blue line.}
\label{mudiff}
\end{figure}

\begin{figure}
\includegraphics[width=0.49\textwidth]{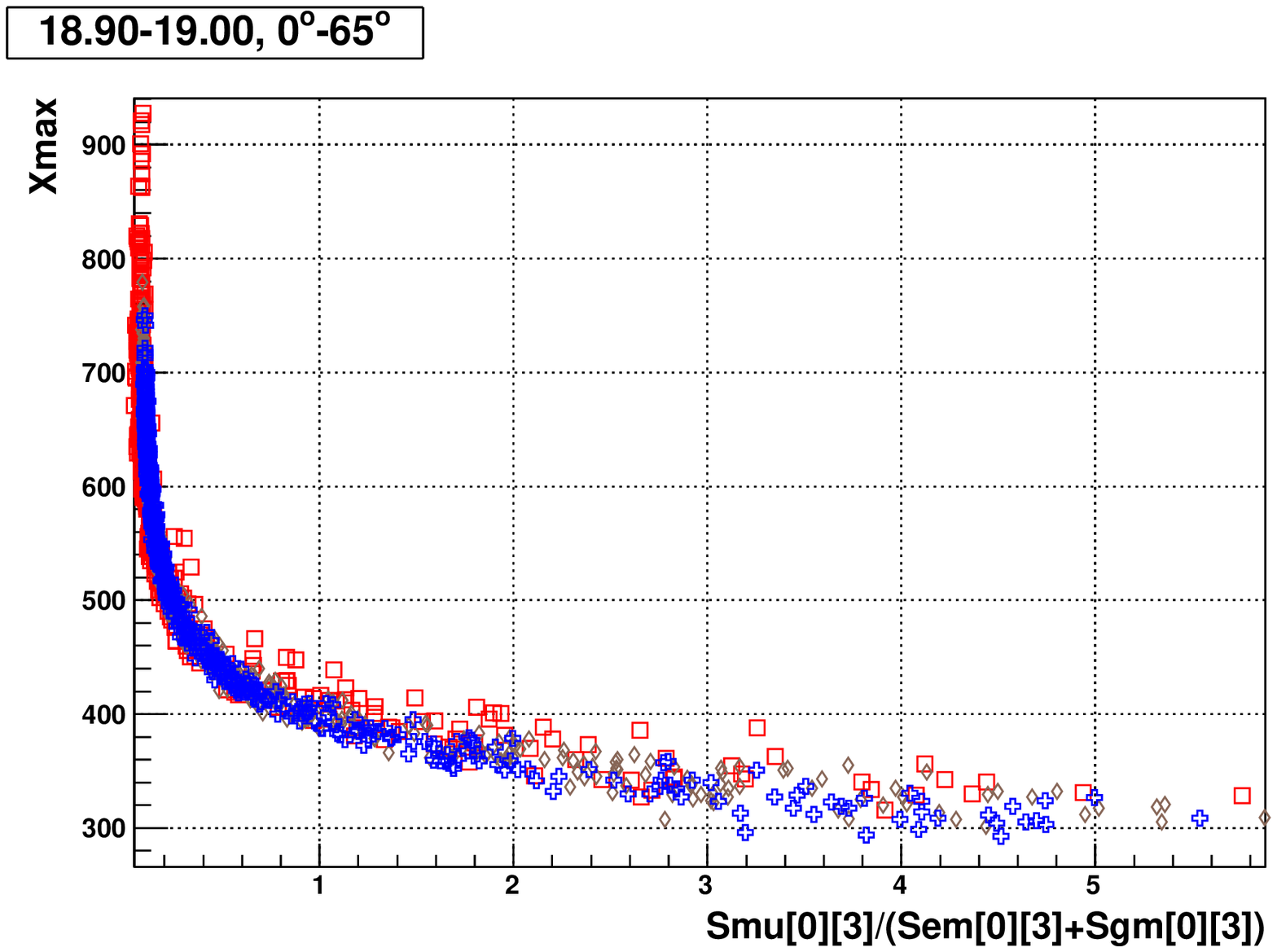}
\includegraphics[width=0.49\textwidth]{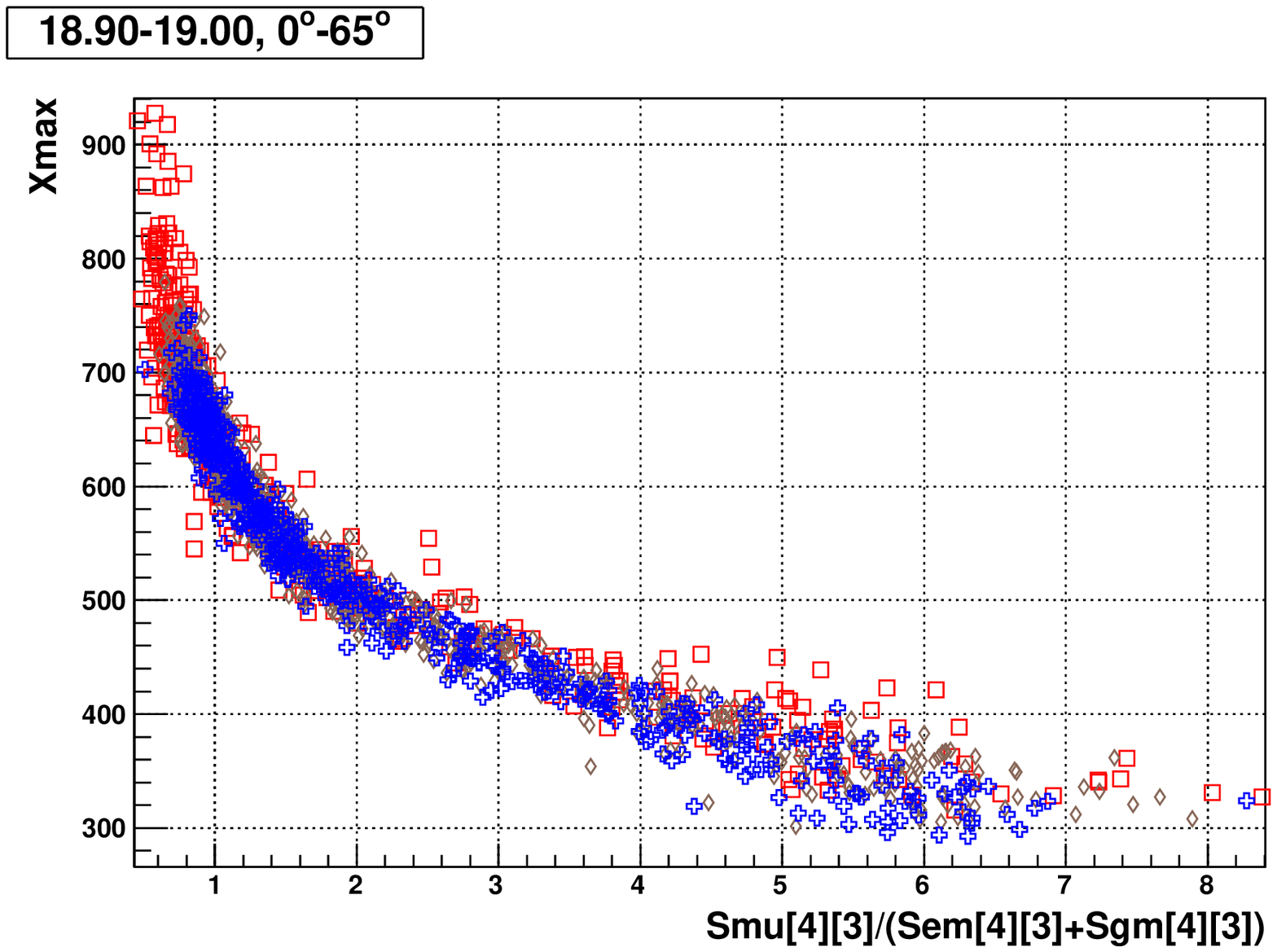}
\caption{Ratio of signals in water Cherenkov tanks \sigrat\ at 200~m
  and 1500~m vs vertical depth of shower maximum \xmaxv\ in
  \logen18.9--19.0 energy bin. Protons~---~red squares,
  oxygen~---~brown diamonds, iron~---~blue crosses}
\label{muemxmax00}
\end{figure}

From different aspects of universality of shower development we will
be interested only in the dependence of electromagnetic and muon
signals on the distance of shower maximum to the ground. We will begin
our consideration from Auger-like experimental setup and consider
signal in water Cherenkov tanks at 1000 meters. In this case one deals
with the situation, where muon contribution to the detector is much
larger compared to the EM one due to higher energy losses of muons in
tanks. The common way to express universality of electromagnetic
signal is to plot it against slant distance to the ground
(Fig.~\ref{DGSlant} and also Figure~1 in~\cite{ave_icrc30_univers}),
showing its quasi-independence on primary particle type. The muon
signal\footnote{Muon signal includes only signal from muons, crossing
  the tank, signal from electromagnetic particles, originating from
  muon decays is included in the electromagnetic signal.} functional
dependence on slant distance to the ground $DG$ is also very similar
for both proton and iron, but there is a shift in the normalization
(Fig.~\ref{DGSlant}). Since iron showers reach \xmax\ earlier than
showers from protons, comparison of set of showers from p and Fe at
equal $DG$ means the comparison between showers with different zenith
angles, but at the same development stage.
Passing to comparison of shower characteristics dependence on vertical
distance to ground \dgv\ reveals a very interesting property (see
Fig.~\ref{DGVert}): in this case the similarity of functional
dependence of muon and EM signals on \dgv\ between p and Fe primaries
is preserved, but now also EM signal normalizations are
different. This happens because one confronts showers, which have the
same vertical distance from \xmaxv\ to the ground, but once again
proton showers are more inclined than iron ones and their EM component
attenuates more while reaching the ground from \xmaxv. The ratio
$S_\mathrm{EM}^\mathrm{Fe}/S_\mathrm{EM}^\mathrm{p}$ turns out to be
almost equal to the $S_\mu^\mathrm{Fe}/S_\mu^\mathrm{p}$ one and this
allows to state the new shower universality property: {\em the ratio
  of the muon signal to the EM one \sigrat\ is the same for all
  showers, reaching the maximum at the same vertical depth \xmaxv,
  independently on the primary particle type, primary energy and
  incident zenith angle} (at the least for the energy and angular
ranges considered here). This property is illustrated in
Fig.~\ref{muemxmax}, where it is shown the dependence of \sigrat\ on
\xmaxv\ for p, O and Fe primaries in four different energy bins. The
functional dependence between \xmaxv\ and \sigrat\ turns out to be
very simple and quasi-universal for all energies and primaries. The
function in the form
\begin{equation}
\label{eq:fit}
\xmaxv=A(S_\mu/S_\mathrm{EM}+a)^b 
\end{equation}
fits well the data and the fit parameters are quite stable across
entire energy range. Having in hand the functional dependence of
\xmaxv\ on (\sigrat) and using $S_\mathrm{tot}=S_\mathrm{EM}+S_\mu$
one easily gets the equation, which allows to obtain muon signal from
shower vertical depth and total signal in tanks:
\begin{equation}
\label{eq:mufit}
S_\mu^\mathrm{fit}=\frac{S_\mathrm{tot}}{1+1/(\left(\xmaxv/A\right)^{1/b}-a)}.
\end{equation}
We calculated the difference between the Monte-Carlo (MC) simulated
muon signal $S_\mu^\mathrm{MC}$ and the muon signal, obtained from the
fit $S_\mu^\mathrm{fit}$, an example distributions of this value are
shown in Fig.~\ref{mudiff}. In Table~\ref{tab:mudiff} we give mean and
RMS values of such distributions for various energy bins, obtained
with the unique set of fit parameters $A=538$, $b=-0.25$ and
$a=-0.22$. It is seen, that the estimates of muon signals are unbiased
with less than 1\% deviation of mean reconstructed muon signal from
the MC one for all 3 primaries and the RMS values are small: 8\% for
protons and around 5\% for oxygen and iron. Certainly, application of
specific coefficients for every energy bin or narrowing of zenith
angle interval, or using of more sophisticated fit functions can even
slightly improve the performance of the method, which anyhow is good
in its simple and universal form. The described universal dependence
of \sigrat\ on \xmaxv\ holds true in the wide interval of distances
and in Fig.~\ref{muemxmax00} we show examples for the distances 200
and 1500 meters from the core, though for distances closer to the core
the function in the form~(\ref{eq:fit}) does not describe accurately
the data in the entire angular range $0^\circ-65^\circ$ and it is
needed or to split it in two parts or to apply more complex
parametrization.

\begin{figure}
\includegraphics[width=0.49\textwidth]{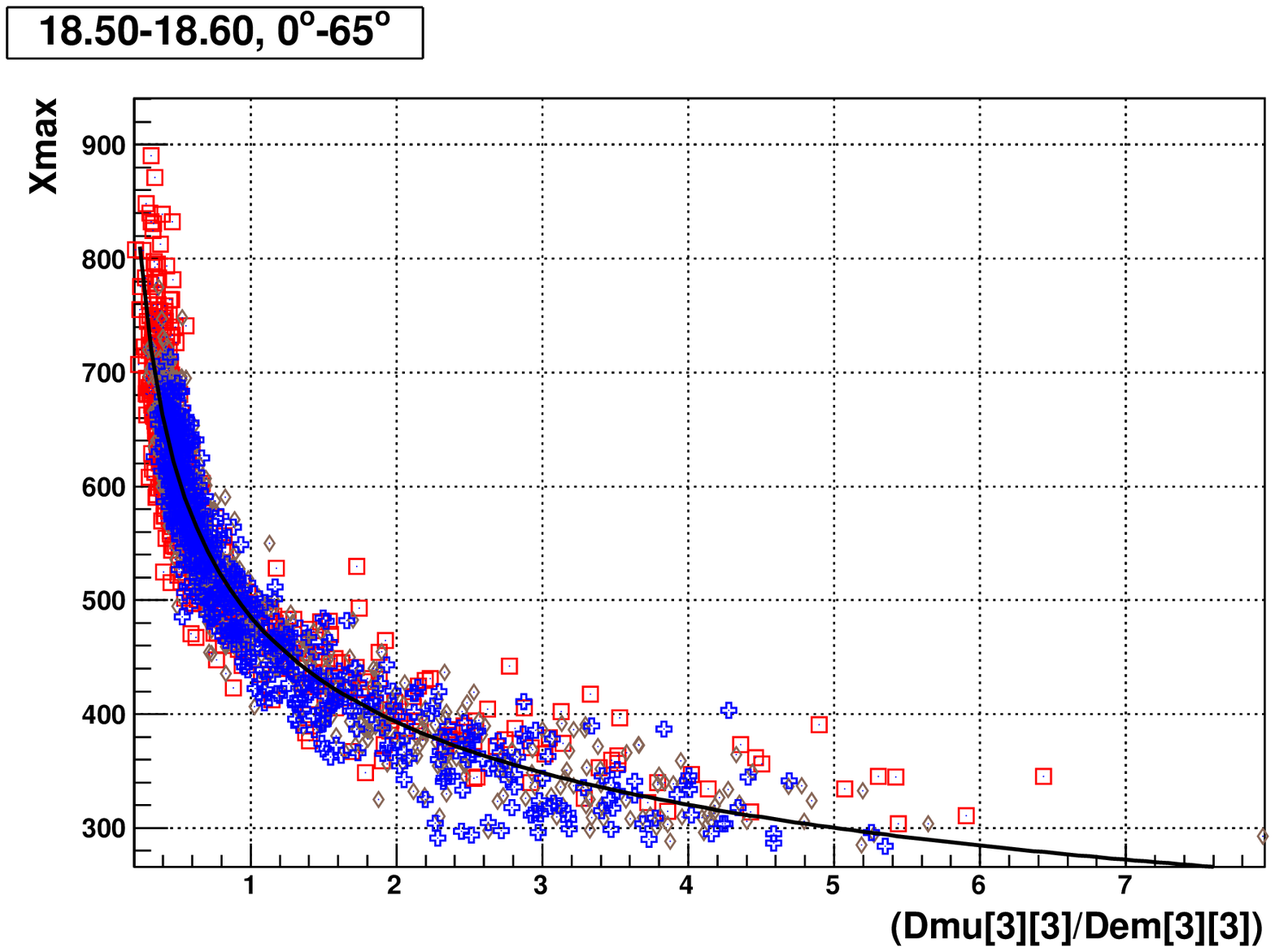}
\includegraphics[width=0.49\textwidth]{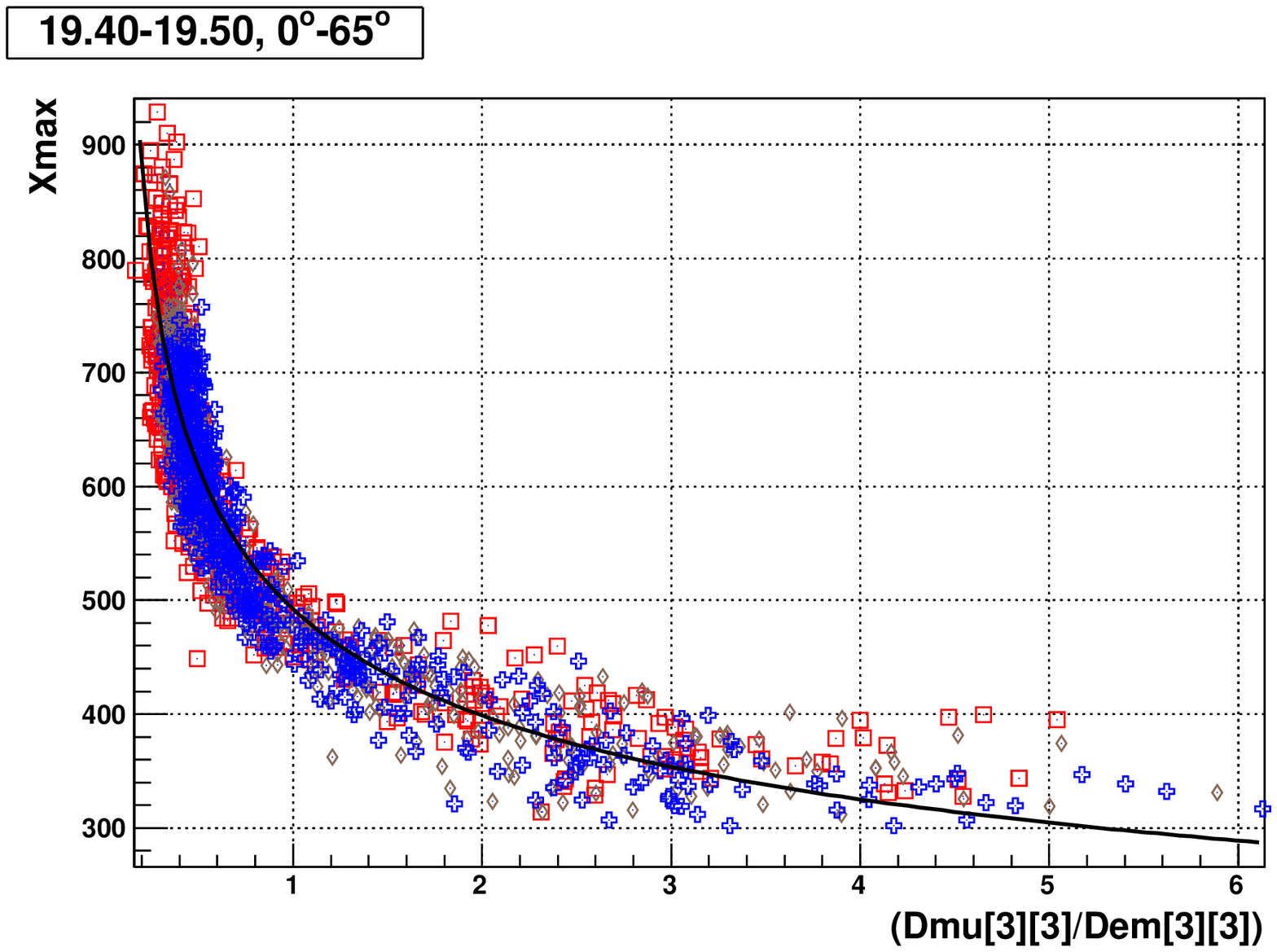}
\includegraphics[width=0.49\textwidth]{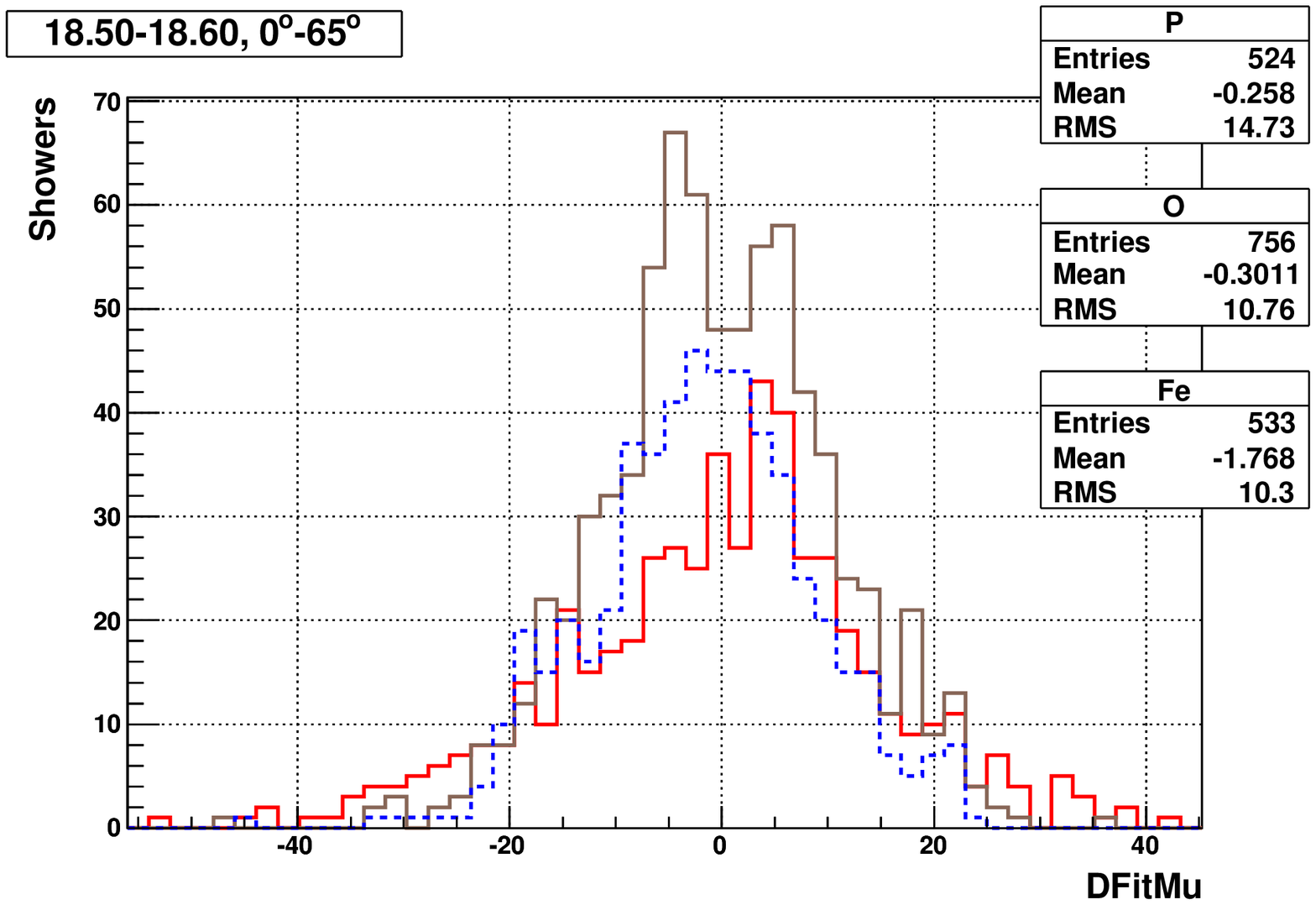}
\includegraphics[width=0.49\textwidth]{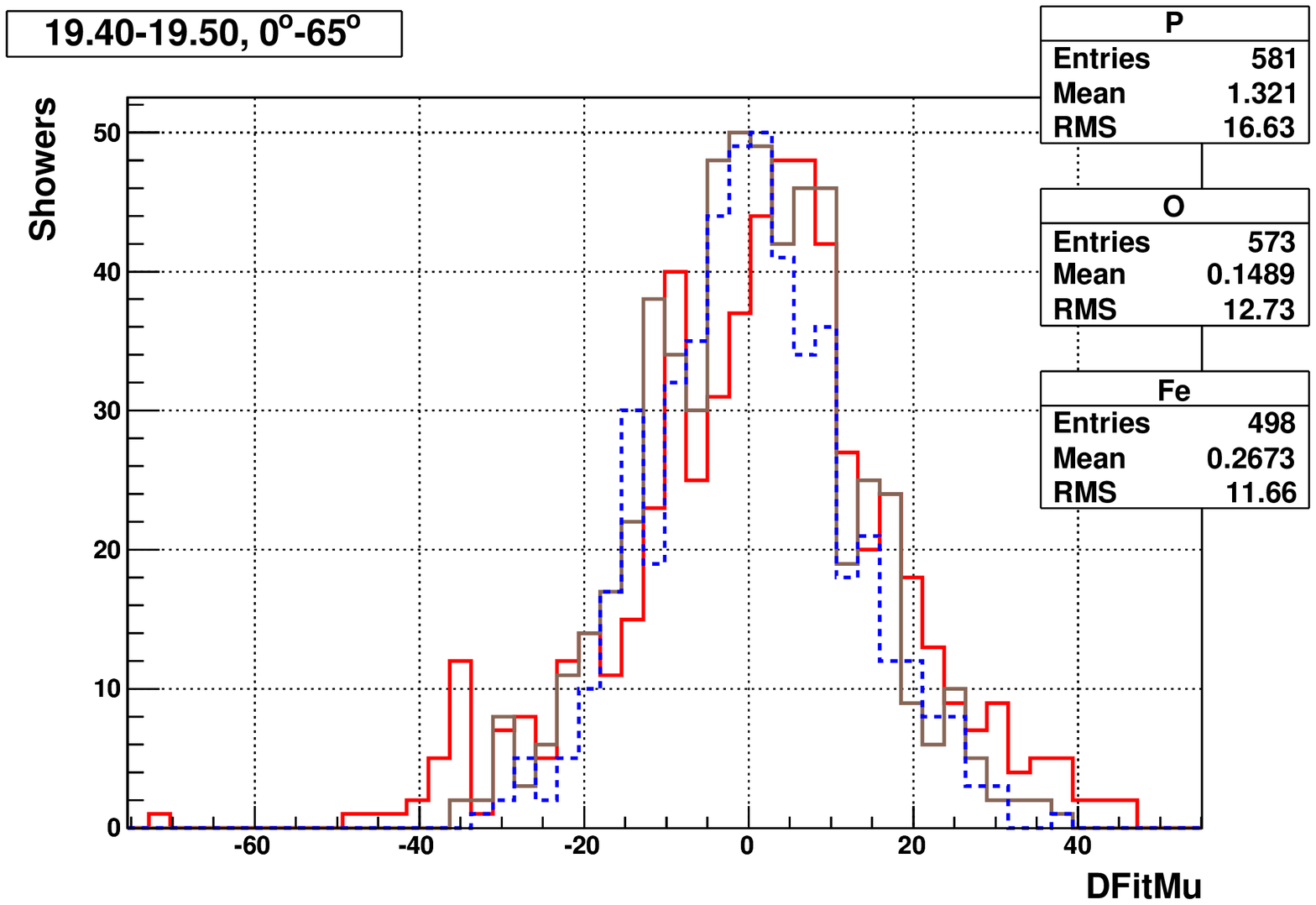}
\caption{Top: ratio of muon density to the electron one at 1000~m
  vs vertical depth of shower maximum \xmaxv\ for two energy bins;
  bottom: distributions of relative difference between MC simulated muon
  density $D_\mu^\mathrm{MC}$ and muon density derived from the fit
  $D_\mu^\mathrm{fit}$ at 1000~m. The data are given for two energy
  bins, protons~---~red,  oxygen~---~brown, iron~---~blue.}
\label{Dmuemxmax}
\end{figure}

The same universality principle holds true in the case of detectors
using scintillators and effectively measuring density of charged
particles~\cite{agasa}. We have performed the reconstruction of muon
densities using the dependence of the ratio (muon density
$D_\mu$)/(electron density $D_\mathrm{e}$) on \xmaxv\ in the
form same to~(\ref{eq:fit}) (see Table~\ref{tab:Dmudiff} and
Fig.~\ref{Dmuemxmax}). It is seen, that when muons and electrons
equally contribute to the detector signal the shower fluctuations play
more important role and the accuracy of parametrization is only within
15\%, though the estimate is still unbiased.

Hence, the new universality property allows to obtain accurate
estimates of the muon signal, which are almost independent on the
primary particle type, primary energy and zenith angle for various
types of ground detectors. Taking in consideration that the shower
universality property was established for different interaction
models~\cite{giller_univers2005,nerling_univers2005,gora_univers2006,ave_icrc30_univers,lipari_univers2008,lafebre_univers2009},
we expect that the proposed approach to muon content derivation is not
specific only to QGSJET~II/Fluka case.

The discovered simple shower universality property in respect to (muon
signal/EM signal) ratio, giving access to the muon shower content, can
open new possibilities as in solution of the global problems, such as
derivation of primary mass composition and understanding of hadronic
interactions properties, so in the number of more particular tasks
(e.g. estimations of primary energy on the basis of pure
electromagnetic signal, primary particle type independent corrections
to the missing energy in experiments using fluorescent light etc.).

\subsection*{Acknowledgements}
We are very grateful to Maximo Ave and Fabian Schmidt for kind
permission to use their GEANT~4 lookup tables in our calculations of
signal from different particles in Auger water tanks.

\begin{table}
\caption{Means and RMS of distributions of relative difference between
  MC simulated muon signals in Cherenkov water tanks
  $S_\mu^\mathrm{MC}$ and muon signals derived from the fit
  $S_\mu^\mathrm{fit}$ at 1000~m (see also Fig.~\ref{mudiff})
  $(S_\mu^\mathrm{MC}-S_\mu^\mathrm{fit})/S_\mu^\mathrm{MC}$, \%,
  calculated with the unique set of parameters for all energy bins:
  $A=538$, $b=-0.25$, $a=-0.22$}
\label{tab:mudiff}
\renewcommand\tabcolsep{8pt}
\renewcommand\arraystretch{1.2}
\centering\begin{tabular}{l|cccccc}
\hline
$\log10(E)$ [eV]  &\multicolumn{2}{c}{proton}&\multicolumn{2}{c}{oxygen}&\multicolumn{2}{c}{iron}\\
              &       Mean &     RMS    &       Mean &     RMS    &       Mean &     RMS   \\
\hline
18.5 -- 18.6  &   -0.1   &      7.4   &     -0.3   &      4.7   &     -0.8    &     4.2   \\
18.6 -- 18.7  &   -0.3   &      7.2   &     -0.2   &      5.0   &     -0.8    &     3.9   \\
18.7 -- 18.8  &    0.1   &      8.1   &     -0.2   &      4.9   &     -0.8    &     4.5   \\
18.8 -- 18.9  &   -0.1   &      8.3   &     -0.3   &      5.2   &     -0.5    &     4.2   \\
18.9 -- 19.0  &     0.5  &       8.3  &      -0.2  &       4.8  &      -0.5   &      4.3  \\
19.0 -- 19.1   &   -0.1   &      7.2   &      0.4   &      5.2   &     -0.3    &     4.2  \\
19.1 -- 19.2   &    0.4   &      8.5   &      0.2   &      5.1   &     -0.1    &     4.6  \\
19.2 -- 19.3   &    0.3   &      7.8   &      0.3   &      5.1   &     -0.1    &     3.9  \\
19.3 -- 19.4   &    0.2   &      7.9   &      0.0   &      5.2   &      0.2    &     4.5  \\
19.4 -- 19.5   &    0.6   &      8.0   &      0.1   &      5.4   &     -0.1    &     4.3  \\
19.5 -- 19.6   &    1.0   &      7.5   &      0.3   &      6.0   &      0.1    &     4.4  \\
19.6 -- 19.7   &    0.2   &      7.7   &     -0.0   &      5.1   &      0.0    &     4.7  \\
19.7 -- 19.8   &    0.7   &      8.1   &      0.3   &      4.9   &     -0.2    &     4.8  \\
19.8 -- 19.9   &    0.2   &      7.2   &      0.5   &      4.9   &      0.4    &     4.9  \\
19.9 -- 20.0   &    1.2   &      9.0   &      0.2   &      5.8   &     -0.1    &     4.9  \\
\hline
18.5 - 20.0  &    0.3    &  7.9      &  0.1      &  5.1     &  -0.2      &  4.4 \\
\hline
\end{tabular}
\end{table}

\begin{table}
\caption{Means and RMS of distributions of relative difference between
  MC simulated muon density $D_\mu^\mathrm{MC}$ and muon density
  derived from the fit $D_\mu^\mathrm{fit}$ at 1000~m (see also
  Fig.~\ref{Dmuemxmax})
  $(D_\mu^\mathrm{MC}-D_\mu^\mathrm{fit})/D_\mu^\mathrm{MC}$, \%,
  calculated with the unique set of parameters for all energy bins
  $A=475$, $b=-0.28$, $a=-0.09$}
\label{tab:Dmudiff}
\renewcommand\tabcolsep{8pt}
\renewcommand\arraystretch{1.2}
\centering\begin{tabular}{l|cccccc}
\hline
$\log10(E)$ [eV]  &\multicolumn{2}{c}{proton}&\multicolumn{2}{c}{oxygen}&\multicolumn{2}{c}{iron}\\
              &       Mean &     RMS    &       Mean &     RMS    &       Mean &     RMS   \\
\hline
18.5 -- 18.6  &     -0.3  &   15  &   -0.3  &   11  &   -1.8  &   10    \\
18.6 -- 18.7  &     -1.0  &   15  &   -0.1  &   11  &   -1.4  &   10    \\
18.7 -- 18.8  &     -0.0  &   16  &   -0.8  &   11  &   -0.9  &   10    \\
18.8 -- 18.9  &     -1.7  &   16  &   -0.2  &   11  &   -0.9  &   10    \\
18.9 -- 19.0  &      1.1  &   17  &   -0.8  &   12  &   -0.4  &   11   \\
19.0 -- 19.1   &    -0.2   &  15   &   0.6   &  12   &  -1.1   &  11  \\
19.1 -- 19.2   &     0.7   &  16   &   0.3   &  12   &  -0.5   &  11  \\
19.2 -- 19.3   &     0.3   &  15   &   0.5   &  13   &  -0.0   &  11  \\
19.3 -- 19.4   &     0.8   &  16   &  -0.9   &  12   &   0.4   &  11  \\
19.4 -- 19.5   &     1.3   &  17   &   0.1   &  13   &   0.3   &  12  \\
19.5 -- 19.6   &     1.3   &  17   &   1.5   &  13   &   0.3   &  12  \\
19.6 -- 19.7   &    -1.2   &  16   &  -0.4   &  13   &   0.7   &  12  \\
19.7 -- 19.8   &     0.9   &  17   &  -0.7   &  14   &  -0.8   &  12  \\
19.8 -- 19.9   &     0.2   &  16   &  -0.1   &  13   &   0.6   &  13  \\
19.9 -- 20.0   &     1.6   &  19   &   0.5   &  14   &   0.4   &  12  \\
\hline
18.5 - 20.0  &       0.3   &  16   &  -0.1   &  12   &  -0.4   &  11 \\
\hline
\end{tabular}
\end{table}

\clearpage

\ifx\undefined\bysame
\newcommand{\bysame}{\leavevmode\hbox to3em{\hrulefill}\,}
\fi

\end{document}